\documentclass[]{aastex631}

	\usepackage{graphicx}	
	\usepackage{amsmath}
	\usepackage{amssymb}	
	\usepackage{longtable}
	\usepackage{threeparttable}
	\usepackage{multirow}
	\usepackage{booktabs}
	\usepackage{array}
    \usepackage[figuresright]{rotating}
    \usepackage[T1]{fontenc}   

	\shorttitle{sGRB $B_P-P_0$}
	\shortauthors{Li et al.}

	\graphicspath{{./}{figures/}}

\begin{document}
		\title{Universal scaling between magnetar field and initial spin period for short gamma ray bursts}

\author{Qin-Mei Li}
\affiliation{Department of Astronomy, School of Physics and Astronomy, Yunnan University, Kunming 650091, China}

\author[0000-0003-0516-404X]{Qi-Bin Sun}	
\affiliation{Department of Astronomy, School of Physics and Astronomy, Yunnan University, Kunming 650091, China}

\author{Sheng-Bang Qian}\email{qiansb@ynu.edu.cn}
\affiliation{Department of Astronomy, School of Physics and Astronomy, Yunnan University, Kunming 650091, China}

\author{Li-Yin Zhu}
\affiliation{Department of Astronomy, School of Physics and Astronomy, Yunnan University, Kunming 650091, China}
\affiliation{Yunnan Observatories, Chinese Academy of Sciences, Kunming 650216, China}
\email{}

\author{Fu-Xing Li}
\affiliation{Department of Astronomy, School of Physics and Astronomy, Yunnan University, Kunming 650091, China}
\email{}

\author{Si-Yuan Zhu}
\affiliation{School of Physics and Astronomy, Sun Yat-Sen University, Zhuhai, 519082, China}
\email{}

\author{Ming Lian}	
\affiliation{Department of Astronomy, School of Physics and Astronomy, Yunnan University, Kunming 650091, China}
\email{}

\author{Jing Li}	
\affiliation{Department of Astronomy, School of Physics and Astronomy, Yunnan University, Kunming 650091, China}
\email{}


\begin{abstract}
	The $B_p$--$P_0$ correlation serves as a critical probe of magnetar engine physics. Although this scaling relation has been firmly established for long gamma-ray bursts (lGRBs), systematic investigations for short GRBs (sGRBs) remain absent, leaving the physical differences between the two populations poorly constrained. Here we analyze 33 Swift sGRBs exhibiting prominent X-ray plateaus from newborn millisecond magnetar spin-down, and derive their initial spin period $P_0$ and polar magnetic field $B_p$. sGRB magnetars span $P_0 \in [1.73,\,18.28]\ \mathrm{ms}$ and $B_p \in [0.06,\,2.82] \times 10^{17}\ \mathrm{G}$ ($\langle B_p \rangle = 7.05 \times 10^{16}\ \mathrm{G}$), significantly more magnetized than lGRB magnetars ($B_p \in [0.39,\,23.08] \times 10^{15}\ \mathrm{G}$; $\langle B_p \rangle = 3.69 \times 10^{15}\ \mathrm{G}$). For the first time, we derive consistent power-law $B_p$–$P_0$ correlations for GRBs: the scaling for sGRBs is $\log B_p = (0.84\pm0.07)\log P_0 + (15.79\pm0.07)$, whose slope is highly consistent with that of lGRBs, $\log B_p = (0.83\pm0.09)\log P_0 + (14.92\pm0.06)$. The near-identical slopes imply a universal magnetar spin-down mechanism, while the vertical offset between intercepts traces divergent progenitor channels. This scaling relation thus offers a new diagnostic to disentangle the formation pathways of GRB. Within the framework of the standard spin-up model, the mass accretion rates of sGRBs ($\dot{M} \sim 1 \times 10^{-1}$ to $3 \times 10^{-1}\,M_\odot\,\mathrm{s}^{-1}$) are substantially higher than those of lGRBs ($\dot{M} \sim 10^{-4}$ to $1 \times 10^{-1}\,M_\odot\,\mathrm{s}^{-1}$). Our work completes the missing $B_p$--$P_0$ statistics for sGRBs, quantitatively unifies their magnetar physics with lGRBs, and provides new observational constraints on the origin diversity of relativistic transients.

\end{abstract}

\keywords{Gamma-ray bursts (629); Magnetars(992); X-ray astronomy(1810)}


\section{Introduction} \label{sec:intro}	

Gamma-ray bursts (GRBs) are among the most energetic astrophysical phenomena in the universe, characterized by intense emissions of high-energy photons. According to the internal shock model \citep{1997ApJ...490...92K,1999PhR...314..575P}, the central engine produces shells with comparable energies but different Lorentz factors $\Gamma$, where collisions between faster and slower shells generate the characteristic pulse profiles observed in most GRBs. The exceptional luminosity of GRBs makes them valuable probes of the distant universe, with GRB 090429B holding the record as the most distant detected burst at $z=9.4$ \citep{2011ApJ...736....7C}. GRBs have long been utilized as tracers of the cosmic star formation rate (SFR) and delayed star formation history \citep{2025ApJ...990L..54L,2026ApJ...998..121L}, and they serve as powerful diagnostic tools to constrain the origins of other transients (eg., extragalaxies X-ray fast trasient, Fast radio burst) \citep{2024ApJ...973L..54C,2026ApJ...997L..15L}. GRBs are traditionally classified into two categories based on their duration $T_{90}$ \citep{1993ApJ...413L.101K}, with long-duration GRBs (lGRBs; $T_{90} > 2\,\text{s}$) widely accepted to result from the collapse of massive stars. This connection is supported by direct observational associations between lGRBs and supernovae \citep{2003ApJ...591L..17S,2004ApJ...609..952Z,2023MNRAS.524.1096L}. sGRBs are predominantly associated with the mergers of compact binary systems, such as binary neutron stars (BNS) or neutron star–black hole (NS–BH) binaries \citep{1986ApJ...308L..43P, 1989Natur.340..126E, 1992ApJ...395L..83N}. The detection of GW170817 and its electromagnetic counterparts—GRB 170817A and the optical transient AT 2017gfo—provided compelling observational evidence that BNS mergers can give rise to sGRBs \citep{2017ApJ...848L..13A}.

GRB radiation could originates from fireball emission from the accretion disk of newborn black holes formed either via binary mergers or failed Type Ib supernovae from collapsing rotating Wolf-Rayet stars \citep{1993ApJ...405..273W}. However, this theroy was challenge, Since the launch of the \textit{Swift} satellite in 2004, its abundant X-ray observational data have provided unprecedented valuable constraints for investigating the central engines of GRB, the existence of central engine activity following at least some GRBs in the form of X-ray flares \citep{2005Natur.438..994B,2006A&A...454..113C}, and internal plateaus with rapid decay at the end of the plateaus \citep{2010MNRAS.409..531R,2013MNRAS.430.1061R}. These observations are difficult to interpret within the framework of a black hole central engine, but are consistent with a rapidly spinning millisecond magnetar as the central engine \citep{2014MNRAS.438..240G,2020ApJ...895...58G}. The magnetar scenario has been invoked to interpret the X-ray afterglow internal plateau emission of both some sGRBs and lGRBs \citep{2011MNRAS.413.2031M,2013MNRAS.430.1061R,2024ApJ...962L..27D,2026ApJ...997L..15L}. These plateaus can be interpreted as a newly born magnetar through magnetic dipole radiation dissipates its rotational energy into electromagnetic output to power GRBs and their multi-wavelength afterglows, the magnetar will collapses into a black hole at the end of the plateau \citep{2014ApJ...780L..21Z}. 

The magnetar model provides a complete theoretical framework to interpret the plateau phase in GRB X-ray afterglows, and enables us to invert key central engine parameters, namely the initial spin period $P_0$ and surface polar magnetic field $B_p$ of newborn magnetars, from observable quantities such as plateau luminosity and plateau duration. Nevertheless, most existing statistical analyses focus solely lGRBs \citep{2020ApJ...903L..24L,2022ApJ...934..125X,2026ApJ..1004L..11Z}. Although \citet{2014ApJ...785...74L} has previously compared the $B_p$–$P_0$ relations of magnetar-powered lGRBs and sGRBs, their analysis only included a small sample of merely 9 sGRBs, lacking sufficient statistical representativeness. Systematic selection and robust parameter constraint of large-scale magnetar candidates originating from sGRBs remain absent, as do comprehensive comparative studies based on ample sGRB samples. To fill these research gaps, we perform uniform fittings to the X-ray afterglow light curves of \textit{Swift} sGRBs and construct a clean sample of magnetar candidates featuring internal plateau signatures. We systematically constrain the distribution of magnetar parameters $B_p$ and $P_0$ for these sGRBs, and further compare our results with published parameter ranges of lGRB magnetar candidates. This allows us to quantitatively characterize the distinct distribution patterns of the two GRB populations across the $B_p$–$P_0$ parameter plane and unravel the fundamental physical dichotomy of their magnetar central engines.

In this work, we construct a sample of 33 sGRBs (22 from \citealp{2015ApJ...805...89L}) to systematically explore magnetar central engine properties and compare the derived $B_p - P_0$ statistical distributions with those of 169 lGRBs, data from \citet{2026ApJ..1004L..11Z}, superluminous supernovae (SLSNe), fast blue optical transients (FBOTs), and Type Ic SNe. This paper is organized as follows. Section~\ref{sec:methods} introduces the sample selection, analysis procedures and magnetar model, Section~\ref{sec:Results} presents our constrained magnetar parameters and $B_p - P_0$ correlations. Section~\ref{sec:DISCUSSION} provides the overall conclusions and discussions. A flat $\Lambda$CDM cosmology with $\Omega_m = 0.3$ and $H_0 = 70\ \mathrm{km\ s^{-1}\ Mpc^{-1}}$ is adopted throughout our analysis.

\section{Sample and methods}\label{sec:methods}
The \textit{Swift} satellite supplies GRB light curves consisting of both prompt emission and X-ray afterglow components. We retrieve all light-curve datasets from the Swift Burst Analyser (\url{https://www.swift.ac.uk/burst_analyser/}; \citealp{2007A&A...469..379E,2009MNRAS.397.1177E,2010A&A...519A.102E}). BAT (15–150 keV) spectral data are extrapolated to the XRT (0.3–10 keV) band under the assumption of a single power-law spectrum. 
Our methodology follows that described in \citet{2015ApJ...805...89L}, who previously identified a sample of 22 sGRBs featuring internal plateaus spanning GRB 050724 to GRB 120521A. Therefore, we systematically analyze a total of 86 sGRBs observed by \textit{Swift} between May 2012 and October 2025. Several sources are discarded for the following reasons: they are too faint for reliable X-ray detection, lack sufficient photon counts to build well-constrained light curves, or exhibit external plateau decay indices shallower than $-2$. Our final clean sample contains 11 sGRBs that display unambiguous internal plateau signatures (see Fig. \ref{fig:2}).

We use smooth broken power law function to fit the light curve in Figure \ref{fig:2}, which takes the form of \citep{2019ApJS..245....1T,2016ApJS..224...20Y}
\begin{equation}
 F = F_0 \left[ \left( \frac{t}{t_b} \right)^{\omega \alpha_1} + \left( \frac{t}{t_b} \right)^{\omega \alpha_2} \right]^{-1/\omega}
 \label{equ:1}
\end{equation}
 where $t_b$ is the break time, $F_b=F_0*2^{-1/w}$ is the corresponding flux at the break time $t_b$, $a_1$ and $a_2$ are the indices before and after the break. The smoothness parameter $\omega$ describes the sharpness of the break. 
 
 The X-ray luminosity at the time of $T_b$ is given by 
 \begin{equation}
  L_b = \frac{4\pi d_L^2 F_b}{(1+z)^{1-\beta}}
  \label{equ:2}
 \end{equation}
 where $z$ is the redshift and the $\beta=\Gamma-1$ is the spectral index. The $(1+z)^{1-\beta}$ is used to perform the K-correction, which converts the luminosity to the 0.3–10 keV range in the rest frame of GRBs. The internal plateau was defined as a plateau followed a decay with indice -2 or steeper than 3 \citep{2015ApJ...805...89L,2026ApJ...997..127H}, the $t^{-2}$ decay is expected by the magnetar dipole spin-down model, and steeper slope is an indication that the emission is powered by the internal dissipation of the magnetar wind. The light curves of these 11 sGRBs with internal plateau are presented
 in Figure \ref{fig:2}, along with the smooth broken power-law fits. The values of parameters and fitting result can be get from Table \ref{tab:1}.

Prior to this collapse event, the supramassive neutron star loses angular momentum via multiple braking torques. Among these dissipation channels, magnetic dipole radiation dominates the spin-down process. Both the characteristic spin-down timescale $\tau$ and the corresponding spin-down luminosity $L_0$ are functions of the initial angular velocity $\Omega_0 = 2\pi/P_0$ and the polar surface magnetic field strength $B_p$. Their analytic formulations, originally derived in \cite{2001ApJ...552L..35Z,2015ApJ...805...89L}:

\begin{align}
	\tau &= \frac{3c^3I}{B_p^2R^6\Omega_0^2} = \frac{3c^3IP_0^2}{4\pi^2B_p^2R^6}= 2.05 \times 10^3\ \mathrm{s}\ \left(I_{45}B_{p,15}^{-2}P_{0,-3}^2R_6^{-6}\right)
	\label{equ:3}
\end{align}
\begin{align}
	L_0 &= \frac{I\Omega_0^2}{2\tau}= 1.0 \times 10^{49}\ \mathrm{erg\ s^{-1}}\ \left(B_{p,15}^2P_{0,-3}^{-4}R_6^6\right)
	\label{equ:4}
\end{align}
where $B_p$ is the dipole magnetic field strength at the pole, we adopt the canonical neutron star parameters $I_{45} = I/(10^{45}\,\mathrm{g}\cdot\mathrm{cm}^2)$ and $R_6 = R/(10^6\,\mathrm{cm})$. The $\tau$ was defined as $\tau=t_b/(1+z)$, the characteristic spin-down luminosity is essentially the plateau luminosity, which may be estimated as $L_0\simeq L_b$.

Accoding this model, we can derive two magnetar parameters invovles initial spin period $P_0$ and the surface polar cap magnetic field $B_p$, which can calculated from the plateau luminosity $L_0$ and Spin-down timescale $\tau$ (equation \ref{equ:3} and \ref{equ:4}, \citealp{2001ApJ...552L..35Z}) 

\begin{equation}
	B_{p,15} = 2.05\,\mathrm{G}\left(I_{45}R_6^{-3}L_{0,49}^{-1/2}\tau_3^{-1}\right)
	\label{equ:5}
\end{equation}

\begin{equation}
	P_{0,-3} = 1.42\,\mathrm{s}\left(I_{45}^{1/2}L_{0,49}^{-1/2}\tau_3^{-1/2}\right)
	\label{equ:6}
\end{equation}

\section{Results}\label{sec:Results}
\begin{figure}
	\centering
	\includegraphics[width=0.4\textwidth]{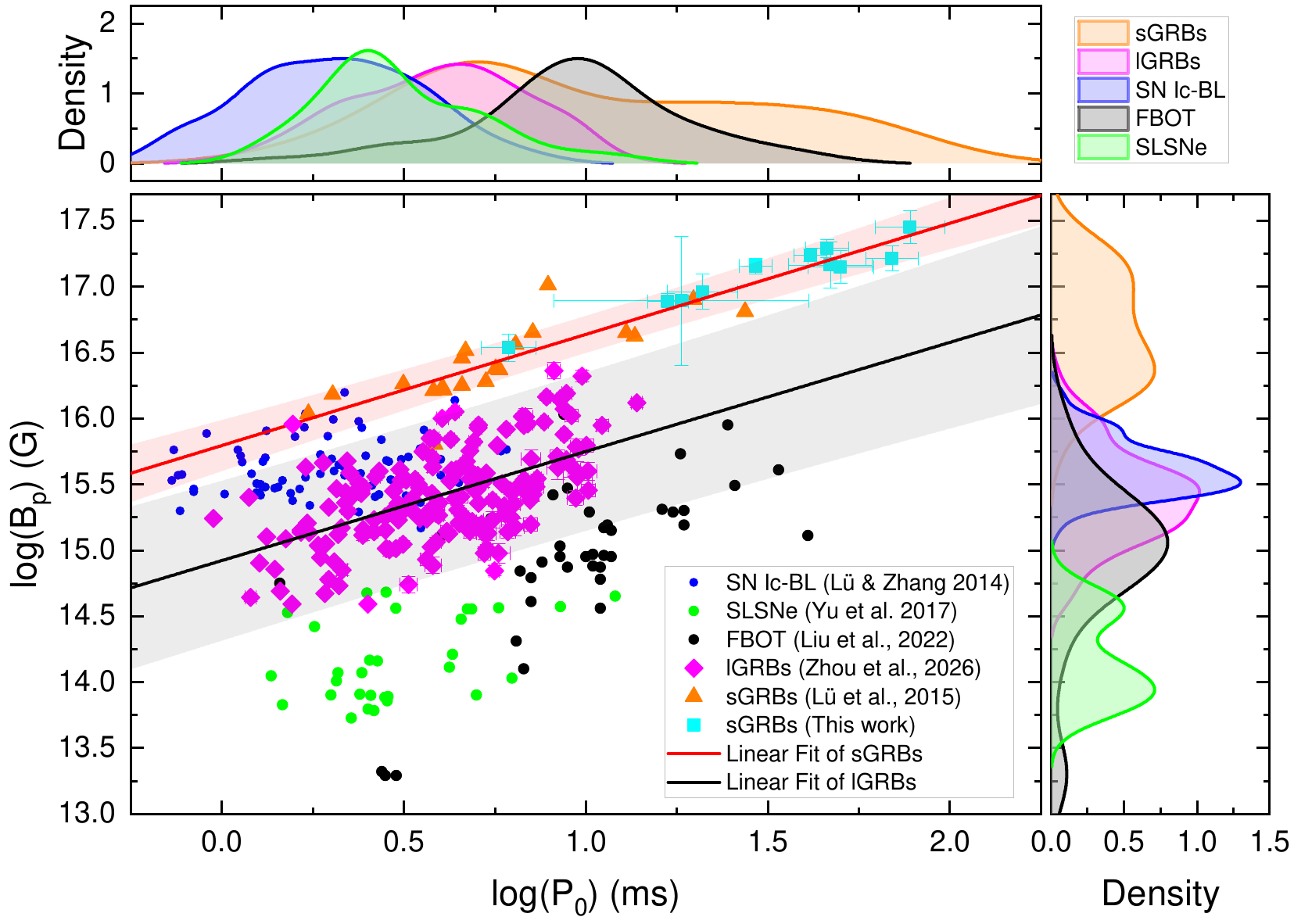}
	\includegraphics[width=0.4\textwidth]{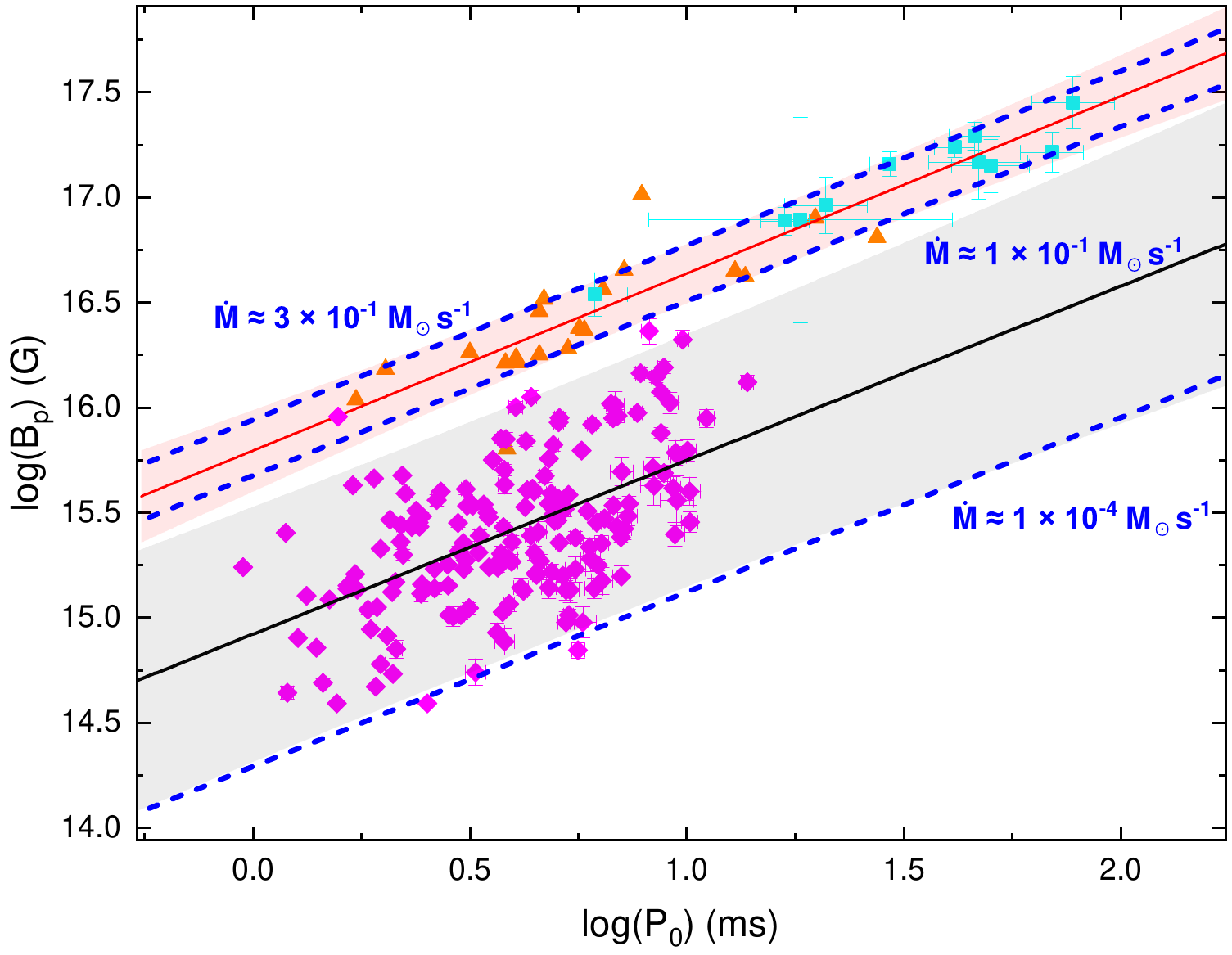}
	\caption{Left: $B_p$–$P_0$ parameter plane of sGRBs and lGRBs powered by magnetar central engines. The two GRB populations show distinct segregation across the magnetic field–initial spin period parameter space; all magnetar parameters are inferred from modeling Swift X-ray afterglow plateaus. Overlaid comparison samples are superluminous supernovae (SLSNe, green points; \citealp{2017ApJ...840...12Y}), lGRBs (violet points; \citealp{2026ApJ..1004L..11Z}), fast blue optical transients (FBOTs, black points; \citealp{2022ApJ...935L..34L}), and broad-line Type Ic supernovae (SNe Ic-BL, blue points; \citealp{2014ApJ...785...74L}). Right: $B_p$–$P_0$ correlations of lGRBs and sGRBs under distinct mass accretion rates.}
	\label{fig:1}
\end{figure}

Figure \ref{fig:1} illustrates the distribution of the parameters $B_p$ and $P_0$. Our results show that sGRBs cover a parameter space of initial spin period $P_0 \in [1.73,\,18.28]\ \mathrm{ms}$ and surface polar magnetic field $B_p \in [0.06,\,2.82] \times 10^{17}\ \mathrm{G}$, with an average $B_p$ value of $7.05 \times 10^{16}\ \mathrm{G}$. For lGRBs, the corresponding parameter ranges are $P_0 \in [0.95,\,13.79]\ \mathrm{ms}$ and $B_p \in [0.39,\,23.08] \times 10^{15}\ \mathrm{G}$, alongside a mean $B_p$ of $3.69 \times 10^{15}\ \mathrm{G}$ \citep{2026ApJ..1004L..11Z}. A direct comparison between sGRBs and lGRBs reveals that sGRBs host magnetars with larger surface polar magnetic fields $B_p$.

Beyond GRBs, magnetars are also widely accepted as plausible central engines that power SLSNe \citep{2017ApJ...840...12Y}, FBOTs \citep{2022ApJ...935L..34L}, and Type SN Ic-BL \citep{2026arXiv260421759Z}. \citet{2017ApJ...840...12Y} carried out parameter analyses for a sample of 31 SLSNe and placed constraints on the intrinsic magnetar properties of these transients. Their work yields progenitor magnetar parameter ranges of $P_0 \in [1.37,\,12.04]\ \mathrm{ms}$ and $B_p \in [0.53,\,4.81] \times 10^{14}\ \mathrm{G}$. Meanwhile, \citet{2022ApJ...935L..34L} studied a set of 40 FBOTs to derive magnetar characteristic parameters, finding that their central magnetars occupy $P_0 \in [1.44,\,40]\ \mathrm{ms}$ and $B_p \in [0.01,\,1.07] \times 10^{16}\ \mathrm{G}$. Separately, \citet{2026arXiv260421759Z} analyzed 80 Type Ic SNe samples and constrained their progenitor magnetar parameters to $P_0 \in [0.73,\,7.73]\ \mathrm{ms}$ and $B_p \in [3.39,\,9.38] \times 10^{15}\ \mathrm{G}$.

Figure \ref{fig:1} further presents the $B_p$–$P_0$ correlation for both sGRBs and lGRBs. We fit the observed samples with a power-law function, whose fitted expressions are given below for sGRBs:
\begin{equation}
	\log B_p = (0.84\pm0.07)\log P_0 + (15.79\pm0.07)
\end{equation}
and for lGRBs \citep{2026ApJ..1004L..11Z}:
\begin{equation}
	\log B_p = (0.83\pm0.09)\log P_0 + (14.92\pm0.06)
\end{equation}
The fitting yields a Pearson correlation coefficient $r=0.93$ with a $P$-value of $2.74 \times 10^{-14}$ for sGRBs, and $r=0.55$ with a $P$-value of $5.4 \times 10^{-15}$ for lGRBs. These values demonstrate a positive correlation between $B_p$ and $P_0$ for both GRB populations. The figure also clearly displays markedly distinct distribution patterns for sGRBs and lGRBs in the $B_p$–$P_0$ parameter plane. The distribution of sGRBs in the \(B_p\)–\(P_0\) parameter plane partially overlaps with that of Type Ic SNe. 

\section{CONCLUSION AND DISCUSSION} \label{sec:DISCUSSION}

In this work, we systematically constrain the $P_0$ and $B_p$ of magnetar central engines in 33 sGRBs, and compare their parameter distributions with those of other magnetar-powered transients---SLSNe, FBOTs, Ic SNe---and 169 lGRBs. The magnetar parameter distributions of sGRBs and lGRBs reveal clearly distinct populations. Quantitatively, sGRB magnetars occupy $P_0 \in [1.73,\,18.28]\ \mathrm{ms}$ and $B_p \in [0.06,\,2.82] \times 10^{17}\ \mathrm{G}$, with a mean $B_p$ of $7.05 \times 10^{16}\ \mathrm{G}$, whereas lGRB magnetars span $P_0 \in [0.95,\,13.79]\ \mathrm{ms}$ and $B_p \in [0.39,\,23.08] \times 10^{15}\ \mathrm{G}$, with a mean $B_p$ of $3.69 \times 10^{15}\ \mathrm{G}$. The systematically stronger polar magnetic fields of sGRB magnetars, which are roughly one order of magnitude larger on average, which is consistent with the findings of \citet{2021MNRAS.508.2505Z}. This confirms that magnetic field strength acts as a critical discriminant separating the progenitor systems of sGRBs and lGRBs, and establishes a fundamental physical dichotomy between the two burst populations.

Power-law fitting of the $B_p$--$P_0$ correlation (Figure~\ref{fig:1}) reveals statistically significant positive correlations in both sGRBs and lGRBs. For sGRBs, the correlation is particularly tight ($Bp\propto P_0^{0.84\pm 0.07}$), with a Pearson coefficient $r = 0.93$ ($P$-value $= 2.74 \times 10^{-14}$), indicating a strong coupling between spin period and magnetic field in the central engine. This constitutes the first systematic detection of the $B_p$--$P_0$ correlation in the sGRB population, providing crucial observational constraints on the central engine properties of compact binary merger products.

The spin-up equilibrium theory for accreting neutron stars in Galactic binaries predicts a natural correlation between magnetic field strength and spin period. When matter accretes onto a magnetized neutron star, the accretion flow is truncated at the Alfv\'en radius $r_m$, where the magnetic pressure balances the ram pressure of the inflowing material \citep{1977ApJ...217..578G, 1991PhR...203....1B}. The neutron star spins up or spins down until it reaches an equilibrium period $P_{\rm eq}$ at which $r_m$ coincides with the corotation radius $r_c = (GM/\Omega^2)^{1/3}$, yielding the fundamental scaling $P_{\rm eq} \propto B^{6/7}\,\dot{M}^{-1/2}$, or equivalently $B \propto P^{7/6}\,\dot{M}^{1/2}$ at fixed $\dot{M}$ \citep{1991PhR...203....1B,2011ApJ...736..108P,2013Ap&SS.346..119P}. Recent studies have further consolidated this universal scaling. Multiple works have derived power-law $B$–$P$ correlations for magnetar central engines of GRBs with X-ray plateaus. \citet{2018ApJ...869..155S} analyzed GRB plateau sources and found $B \propto P^{0.83 \pm 0.17}$. Under the assumption that X-ray plateaus originate from magnetar isotropic winds, \citet{2020ApJ...903L..24L} obtained $B \propto P^{1.13^{+0.11}_{-0.09}}$, while \citet{2022ApJ...934..125X} reported a tight scaling $B \propto P^{1.14}$. \citet{2026ApJ..1004L..11Z} studied a sample of 169 lGRBs and derived $B_p \propto P_0^{0.83 \pm 0.09}$. Additionally, \citet{2026ApJ...997..173L} modeled magnetic and spin evolution for newborn magnetars and yielded $B_p \propto P_0^{1.30 \pm 0.16}$. All these measured slopes are consistent with the theoretical $7/6$ scaling within their respective uncertainties. These converging results from independent samples confirm that the slope of the $B_p$--$P_0$ relation is a universal property of magnetar central engines, governed by the equilibrium between the Alfv\'en and corotation radii during fallback accretion \citep{2023ApJ...949L..32D}.

The most significant finding of this work is that while sGRBs ($\log B_p = (0.84\pm0.07)\log P_0 + (15.79\pm0.07)$) and lGRBs ($\log B_p = (0.83\pm0.09)\log P_0 + (14.92\pm0.06)$) share a consistent power-law slope in the $B_p$--$P_0$ plane, their intercepts differ significantly, with sGRBs exhibiting a systematically higher normalization. This result carries profound physical implications: the slope $B_p \propto P_0^{7/6}$ is set by the equilibrium condition $r_m = r_c$ and thus represents a universal property of magnetar--accretion disk interaction, independent of progenitor type; in contrast, the intercept encodes progenitor-specific accretion conditions, scaling as $\dot{M}^{1/2}$ at fixed $P_0$ \citep{1991PhR...203....1B, 2018ApJ...869..155S}. The higher intercept of sGRBs therefore provides direct observational evidence that compact binary mergers and massive star core collapses produce fundamentally different environments around the newborn magnetar.

Following \citep{1991PhR...203....1B,2013Ap&SS.346..119P,2018ApJ...869..155S}, we express this theoretical relation quantitatively as
\begin{equation}
	\begin{aligned}
		\frac{B_p}{10^{14}\ \mathrm{G}} &\approx 15   \left(\frac{P_0}{1\ \mathrm{ms}}\right)^{7/6} \left(\frac{M}{1.4\,M_\odot}\right)^{5/6} \\
		&\quad \times \left(\frac{\dot{M}}{0.01\,M_\odot\ \mathrm{s}^{-1}}\right)^{1/2} \left(\frac{R}{12\ \mathrm{km}}\right)^{-3}.
	\end{aligned}
	\label{eq:bp_p0_scaling}
\end{equation}
Adopting the same assumptions as \citet{2018ApJ...869..155S} ($M = 1.4\,M_\odot$, $R = 12\ \mathrm{km}$) and our fitted slope of $0.83$, we derive the accretion rates ranges: lGRBs span $\dot{M} \sim 10^{-4}$ to $1 \times 10^{-1}\,M_\odot\,\mathrm{s}^{-1}$ (consistent with \citealp{2018ApJ...869..155S,2020ApJ...903L..24L}), while sGRBs occupy a significantly higher accretion rates of $\dot{M} \sim 1 \times 10^{-1}$ to $3 \times 10^{-1}\,M_\odot\,\mathrm{s}^{-1}$. The standard spin-up model predicts $B_p \propto \dot{M}^{1/2}$ at fixed $P_0$ (Equation~\ref{eq:bp_p0_scaling}), implying that lower accretion rates correspond weaker equilibrium magnetic fields. Since sGRB magnetars exhibit systematically stronger fields, they must correspond to larger accretion rates (see Figure \ref{fig:1} (b)).

In addition to the clear separation between sGRBs and lGRBs on the $B_p$–$P_0$ plane, the $B_p$–$P_0$ distributions of sGRBs also show distinct from other magnetar-powered transients originating from massive stellar core collapse, such as SLSNe \citep{2017ApJ...840...12Y} and FBOTs \citep{2022ApJ...935L..34L}. \citet{2022ApJ...935L..34L} identified a weak correlation between $P_0$ and $B_p$ across lGRBs, FBOTs and SLSNe. For magnetars with identical initial spin periods $P_0$, GRB-associated magnetars statistically possess stronger magnetic fields compared with those of SLSNe and FBOTs. This conclusion is further supported by \citet{2026ApJ..1004L..11Z}, who performed a comparative study of 169 magnetar-driven lGRBs and SLSNe. Their analysis confirms systematically higher magnetic field strengths in GRB magnetars, which points to fundamental distinctions between their progenitor systems or stellar collapse environments. A further critical result is the partial overlap of the $B_p$–$P_0$ parameter space between sGRBs and SNe Ic-BL \citep{2014ApJ...785...74L}. Conventionally, sGRBs are widely interpreted as products of compact binary mergers, our magnetar parameter measurements reveal that a fraction of sGRBs possess progenitor properties nearly identical to those of SNe Ic-BL. The well-studied prototype GRB~200826A, a short burst firmly linked to a Type~Ic supernova, offers direct observational support for this mixed-origin picture \citep{2021NatAs...5..917A, 2022ApJ...932....1R}. This association verifies that massive star core collapse can generate sGRB emission, which naturally explains the overlapping parameter region we recover in our sample. Taken together, our $B_p$–$P_0$ statistics prove that the sGRB population cannot be solely explained by compact binary mergers; instead, sGRBs form via diverse progenitor channels that also include massive stellar core collapse.

These findings deepen our understanding of the intrinsic properties of magnetar central engines and the progenitor physics of GRBs. The quantified population differences and the intrinsic $B_p$--$P_0$ coupling established in this work provide valuable constraints for future numerical simulations and theoretical modeling of GRB progenitor evolution. Larger samples and multi-wavelength joint analyses will help unravel the detailed physics governing magnetar spin--field coupling and contribute to a unified theoretical framework for the diverse class of magnetar-powered astrophysical transients.

\section*{Acknowledgements}
 This work was supported by the Scientific Research Foundation of the Education Department of Yunnan Province (2026Y0165), the Graduate Research Innovation Foundation of Yunnan University (KC-252511615), the National Natural Science Foundation of China (grant Nos. 12503040, 11933008 and 12303040), National Key R\&D Program of China (grant No. 2022YFE0116800), Yunnan Fundamental Research Projects (grant NOs. 202501AS070055, 202503AP140013, 202201AT070092 and 202401AT070143), the Postdoctoral Fellowship Program of CPSF under Grant Number GZC20252095, the China Postdoctoral Science Foundation under Grant Number 2025M773194, Caiyun Postdoctoral Program in Yu nnan Province of China (grant No. C615300504124). This work made use of data supplied by the UK Swift Science Data Centre at the University of Leicester.

\clearpage
\newpage

\renewcommand\thefigure{2}
\begin{figure}
	\centering
	\gridline{
		\fig{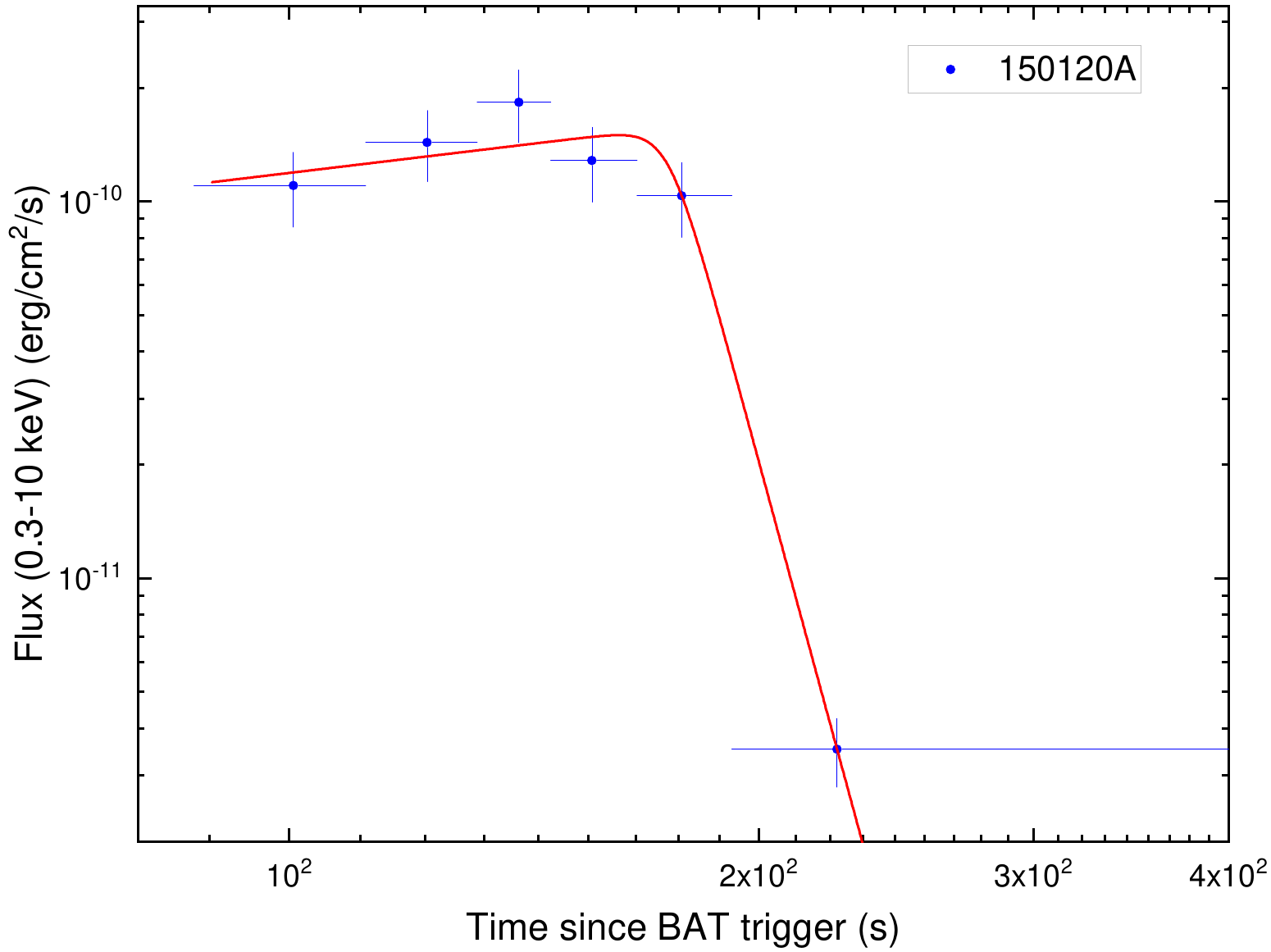}{0.3\textwidth}{}
		\hspace{0.02\textwidth}
		\fig{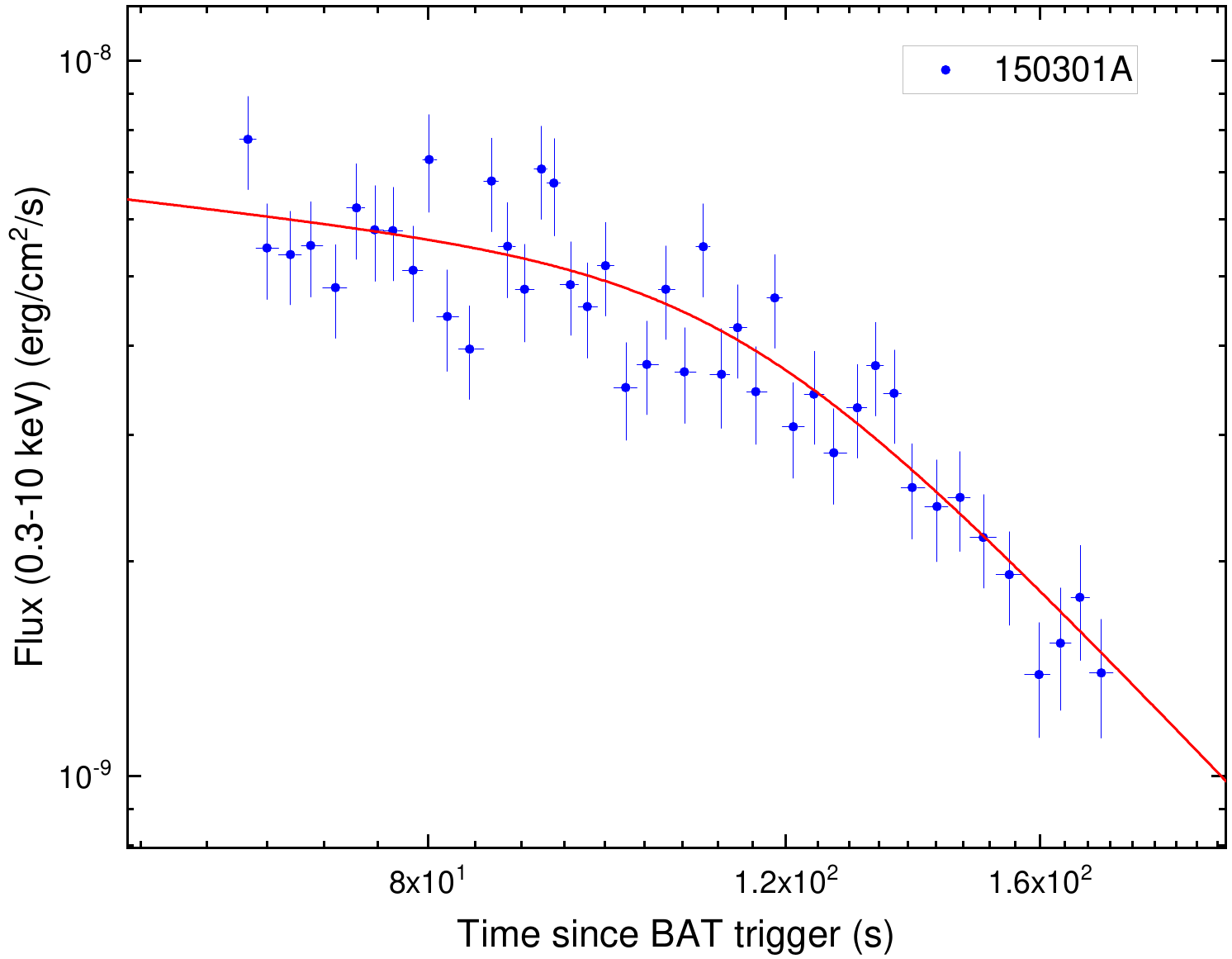}{0.3\textwidth}{}
		\hspace{0.02\textwidth}
		\fig{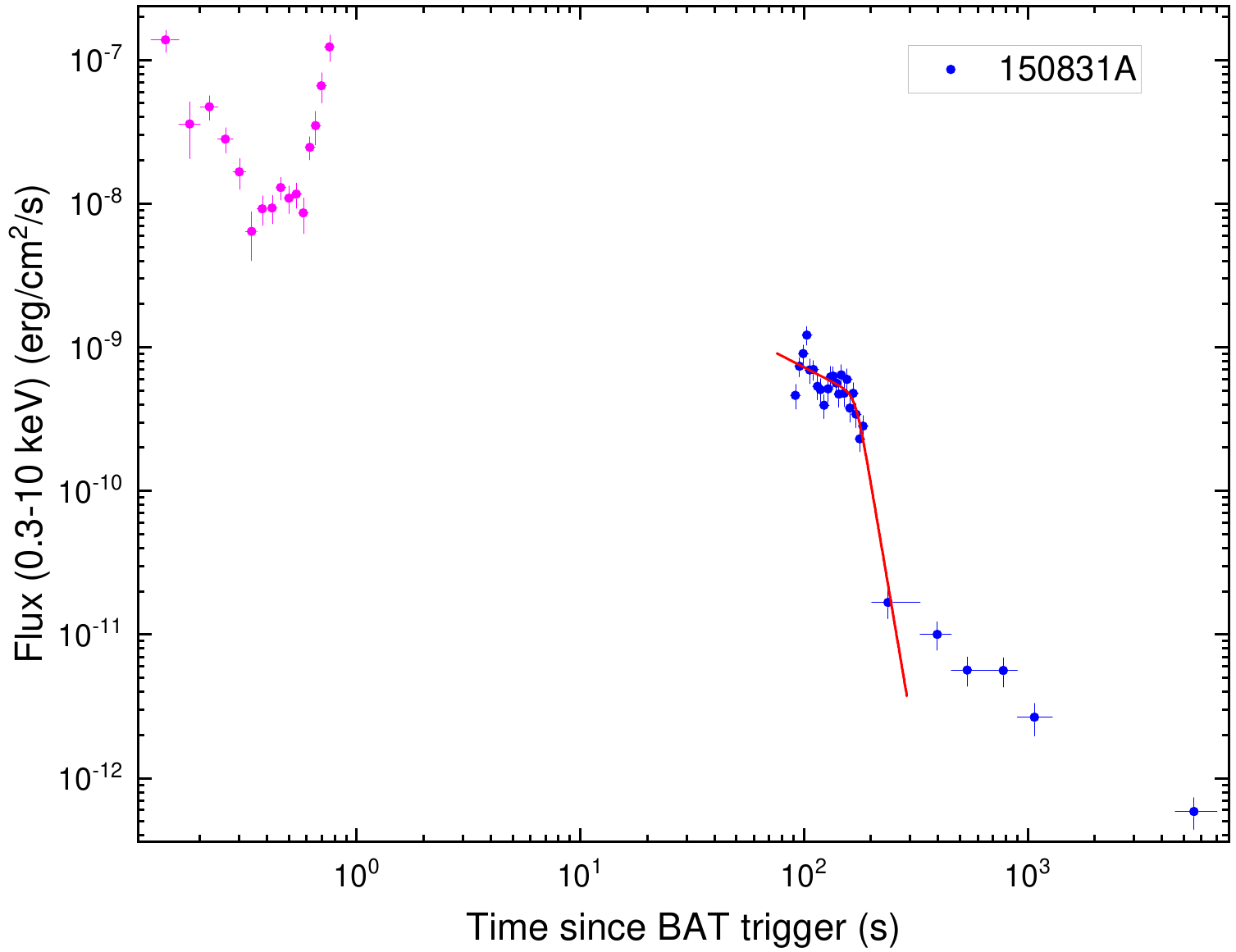}{0.3\textwidth}{}
	}
	\gridline{
		\fig{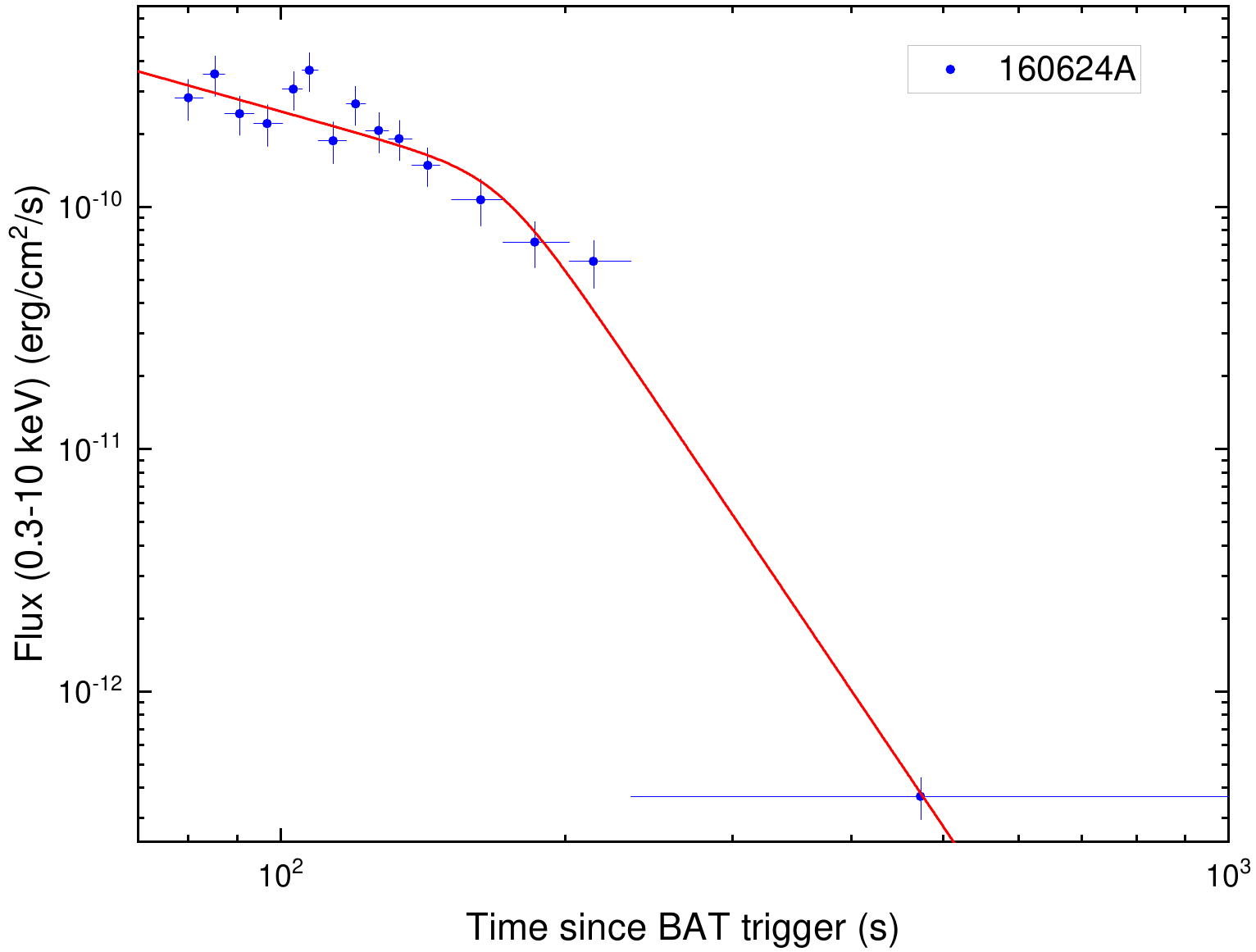}{0.3\textwidth}{}
		\hspace{0.02\textwidth}
		\fig{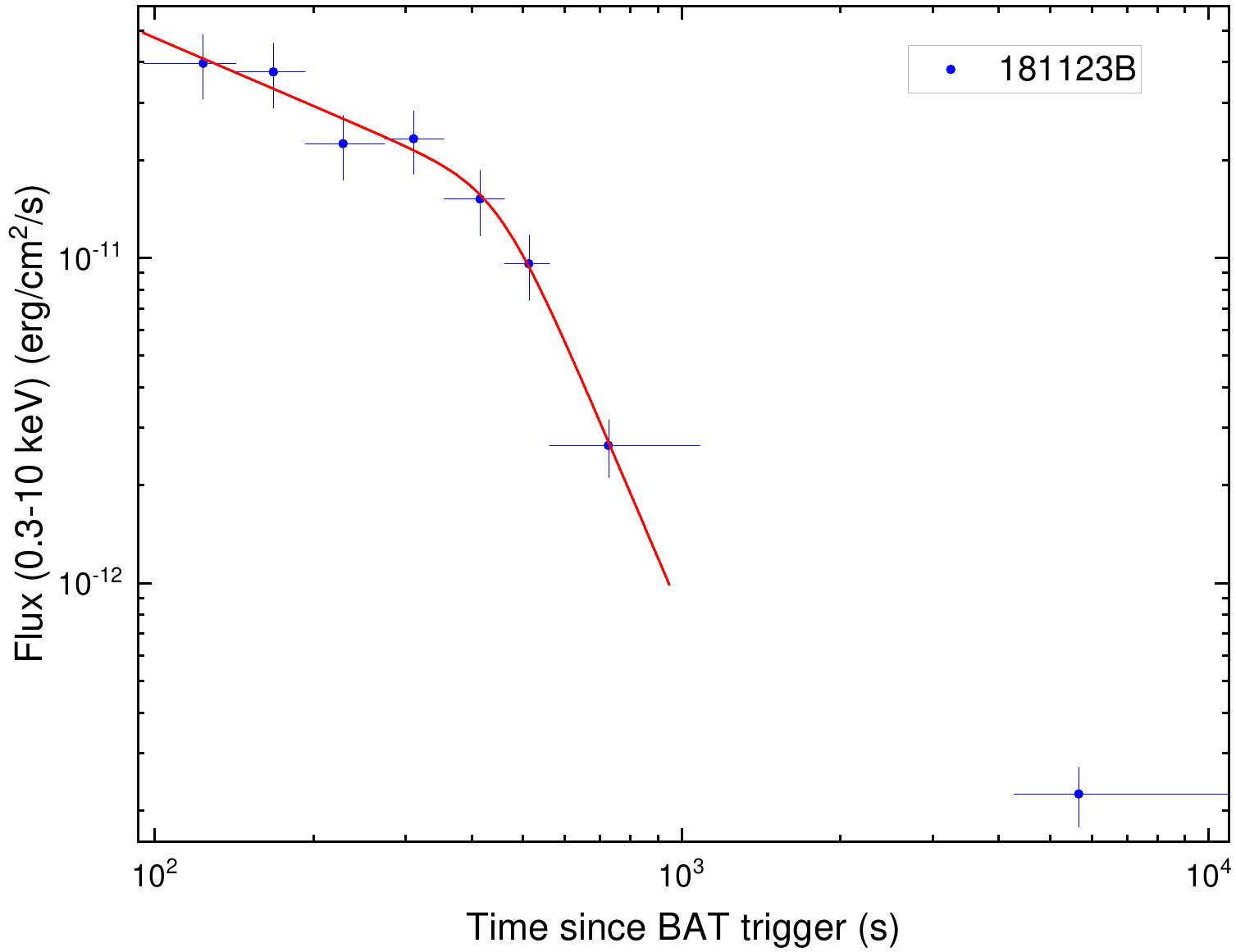}{0.3\textwidth}{}
		\hspace{0.02\textwidth}
		\fig{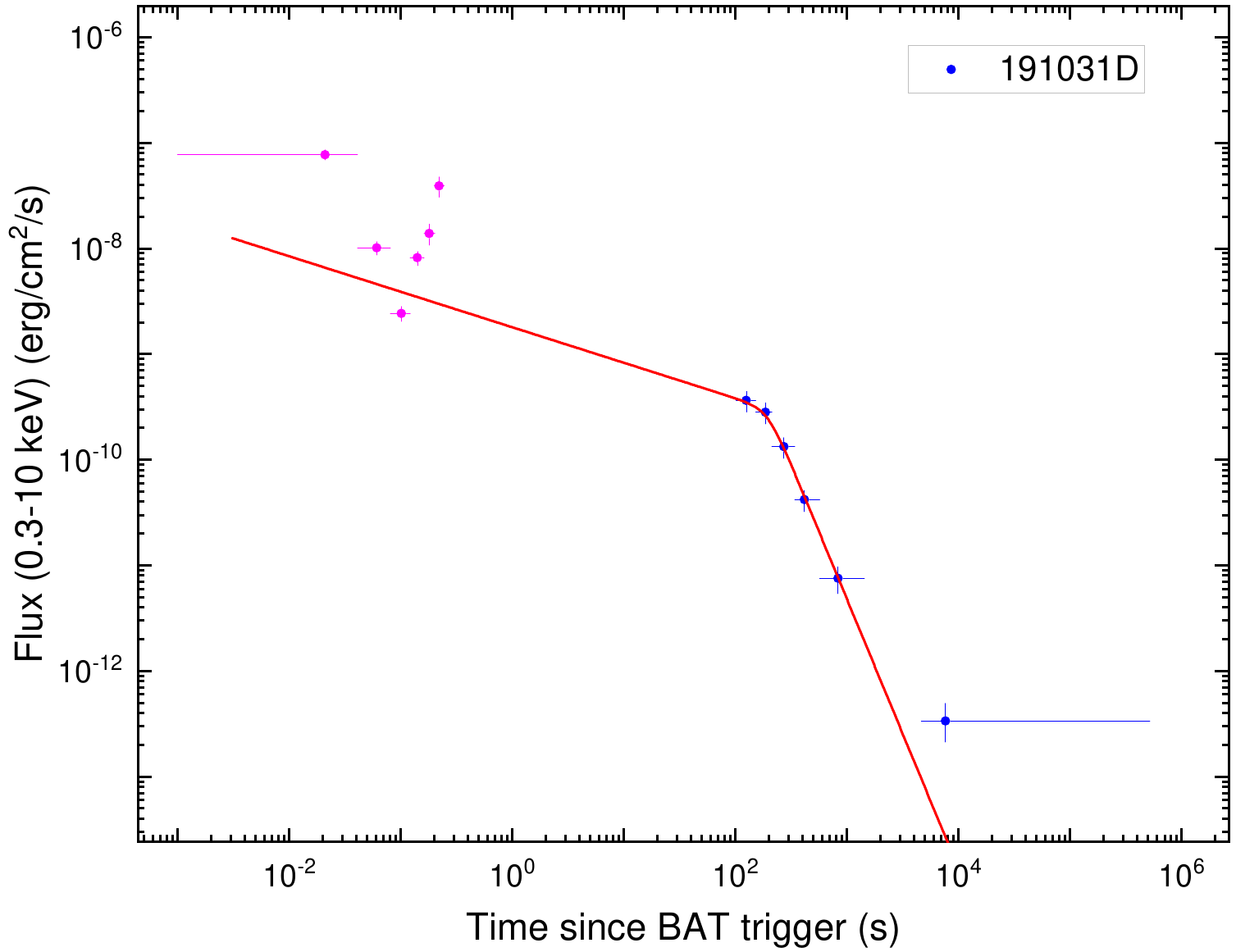}{0.3\textwidth}{}
	}
	\gridline{
		\fig{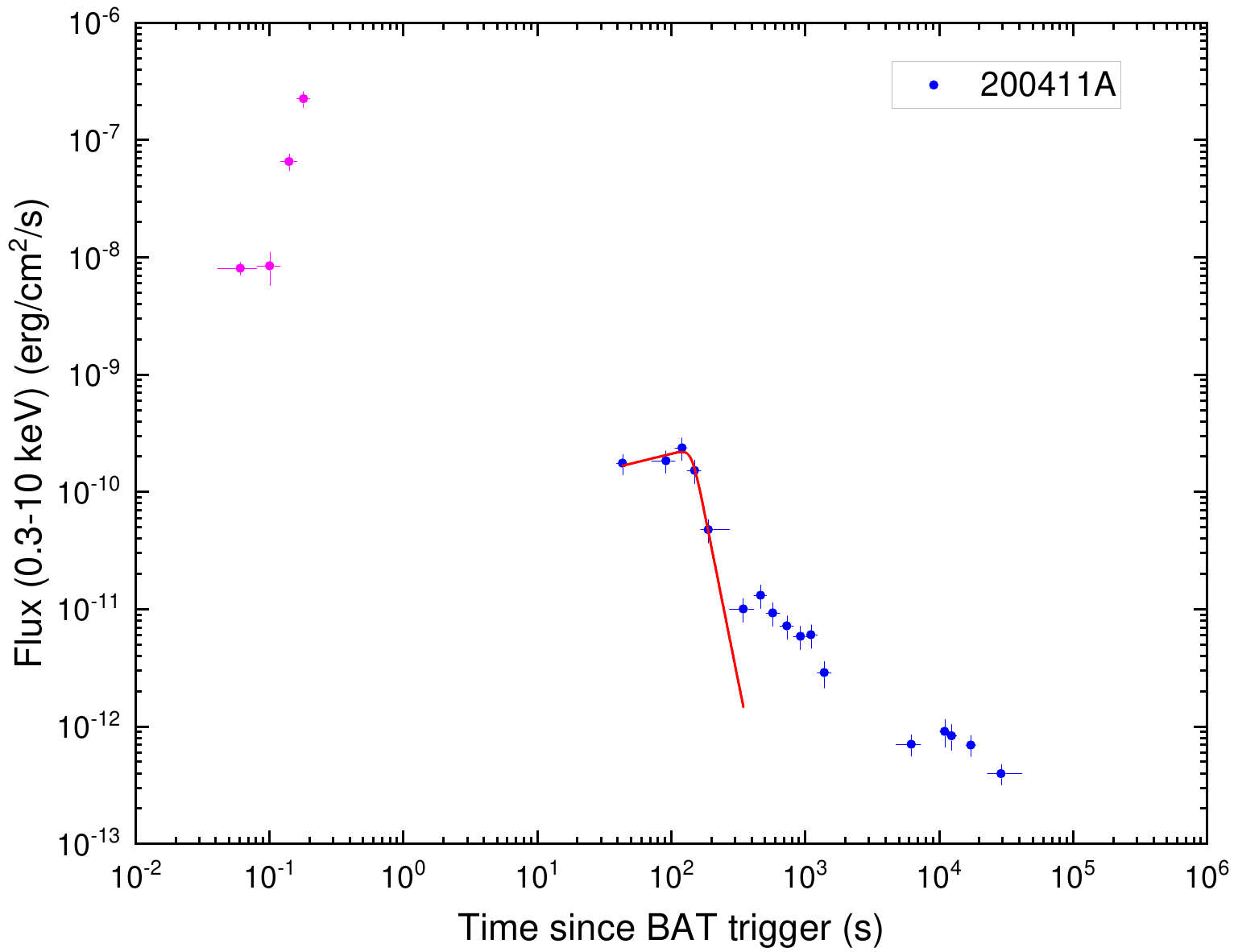}{0.3\textwidth}{}
		\hspace{0.02\textwidth}
		\fig{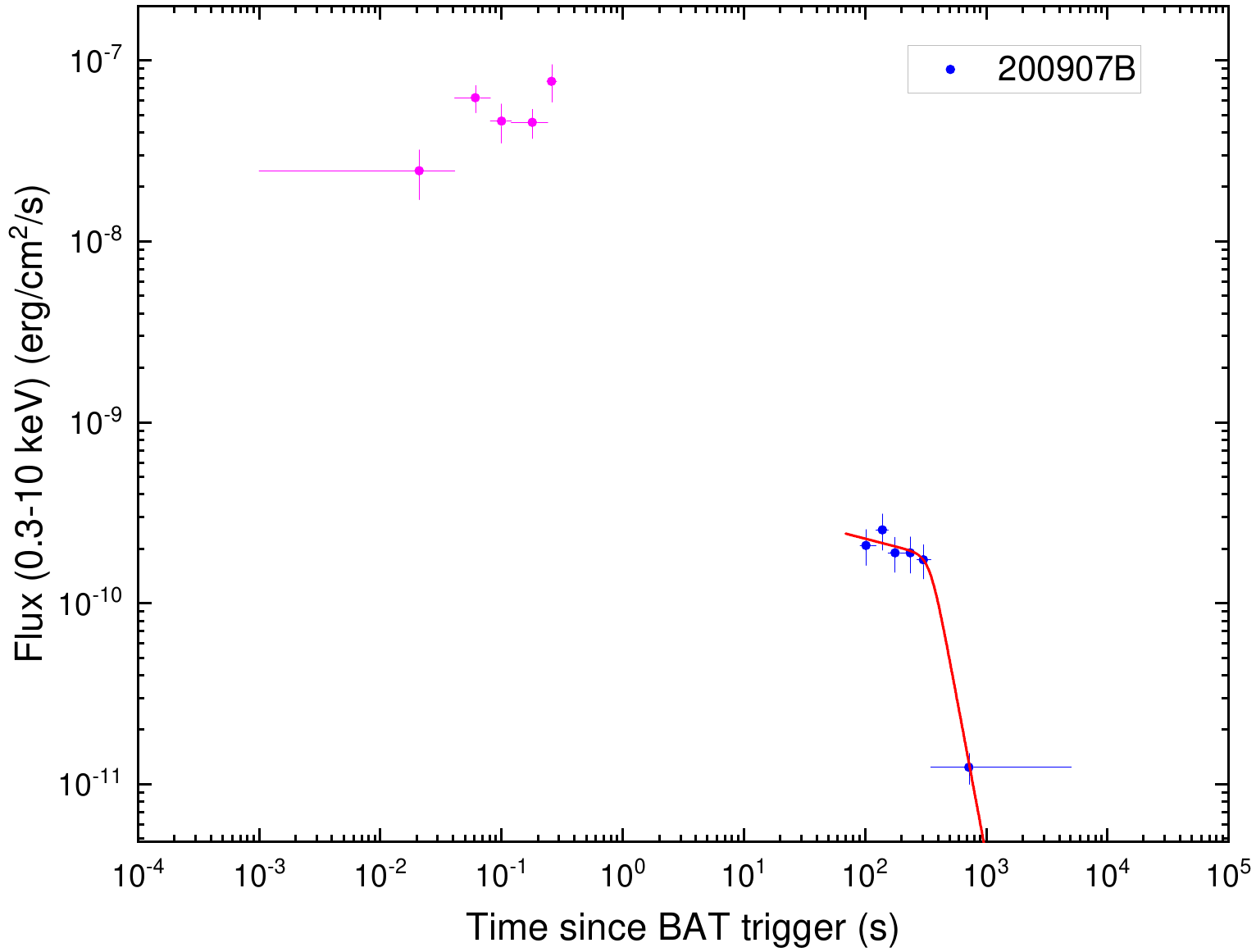}{0.3\textwidth}{}
		\hspace{0.02\textwidth}
		\fig{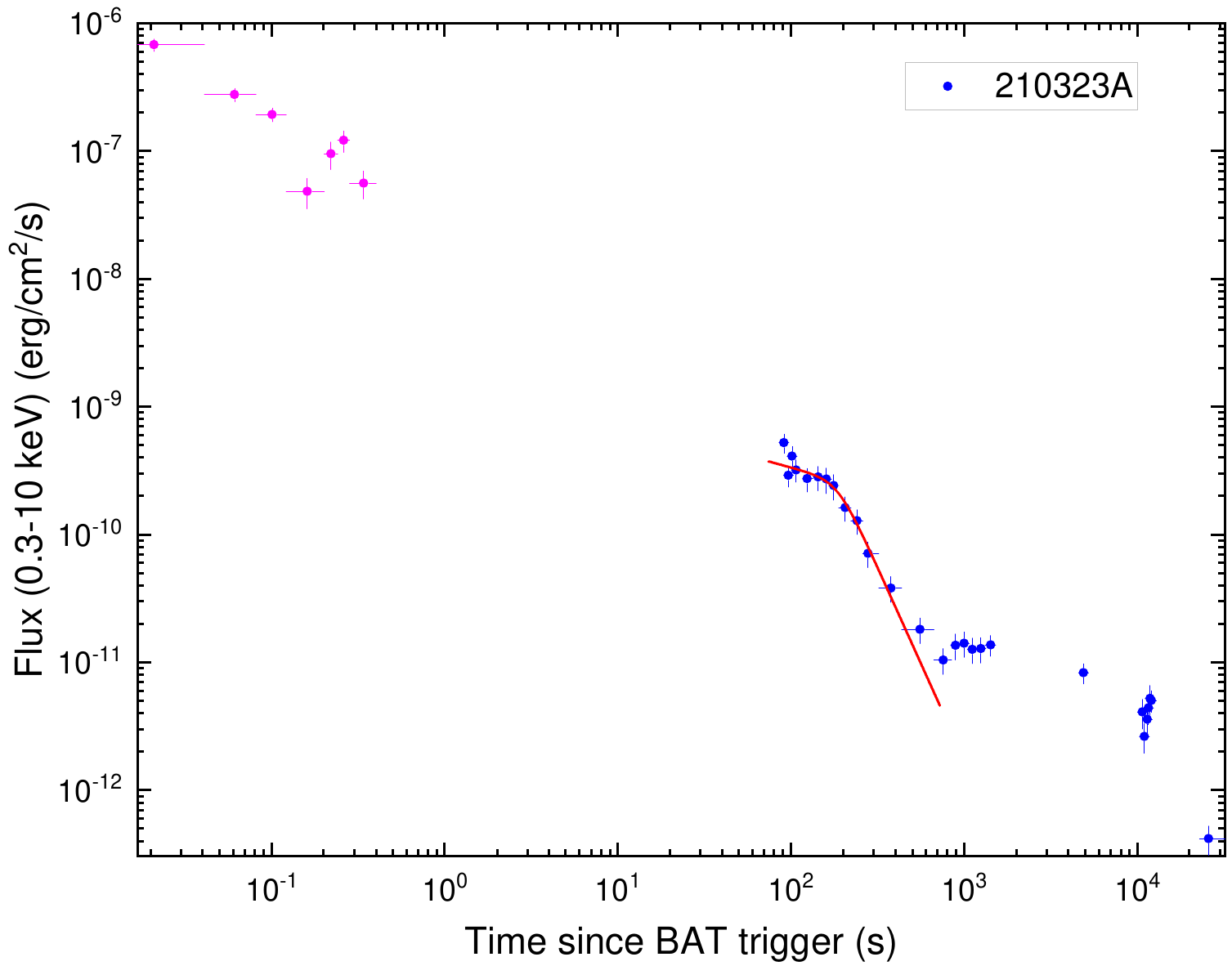}{0.3\textwidth}{}
	}
	\gridline{
		\fig{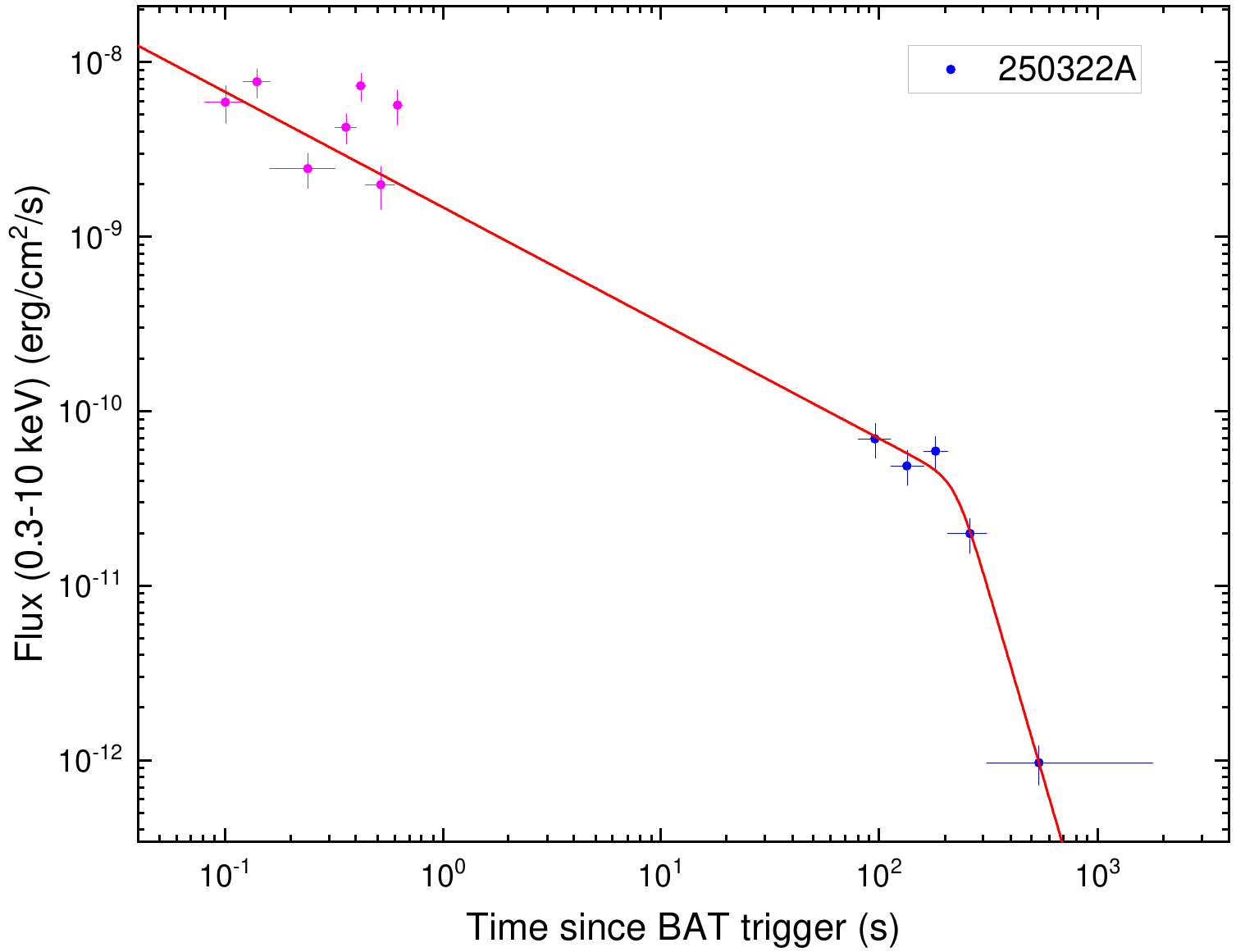}{0.3\textwidth}{}
		\hspace{0.02\textwidth}
		\fig{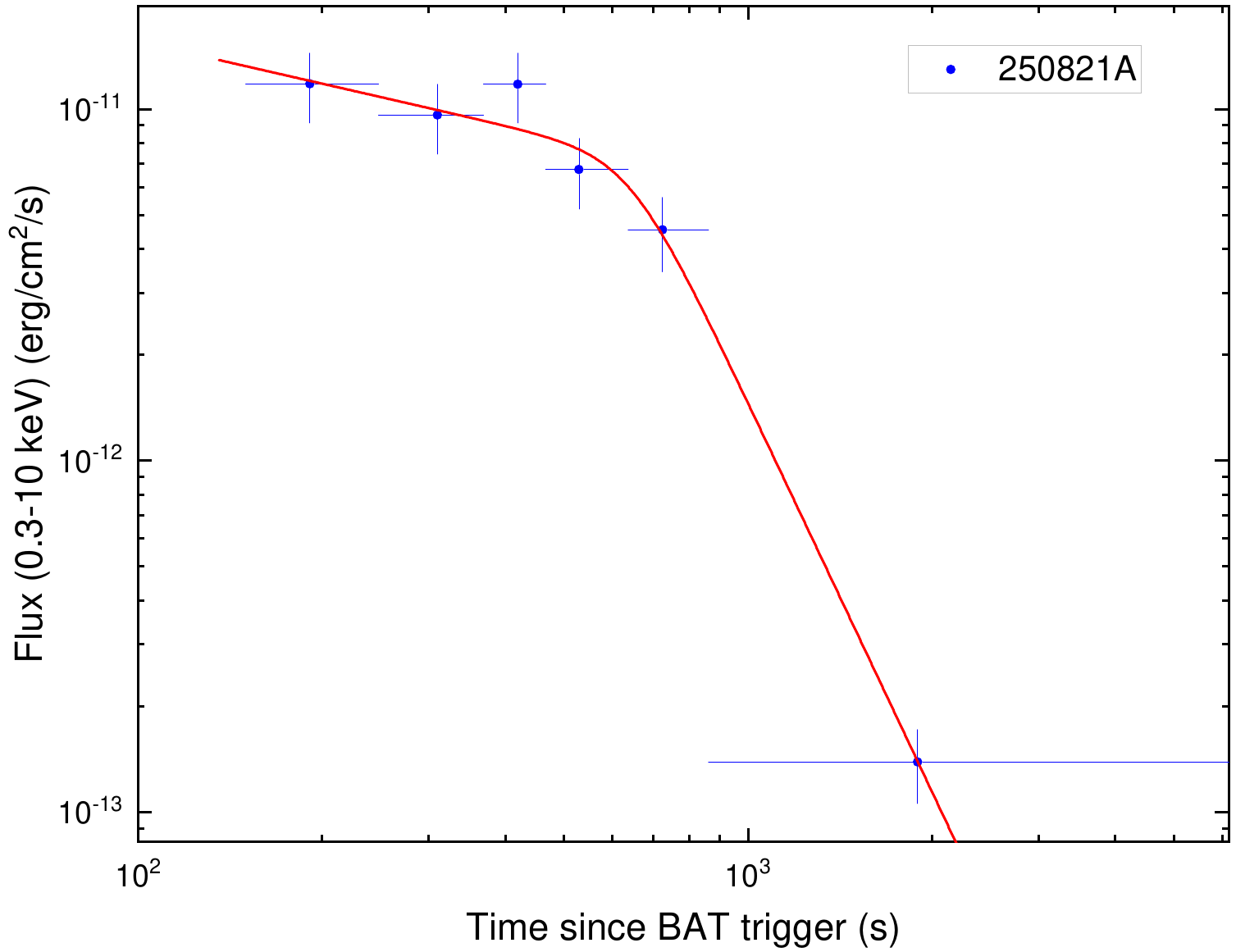}{0.3\textwidth}{}
		\hspace{0.02\textwidth}
	}
	\caption{X-ray light curves of the 11 sGRB sample. Violet points show BAT (15–150 keV) data extrapolated to the XRT (0.3–10 keV) band, and blue points show raw XRT (0.3–10 keV) data. Red lines indicate the best-fit broken power-law models for each light curve.}
	\label{fig:2}
\end{figure}

\newpage

\clearpage

\setlength{\tabcolsep}{1mm}{
	\renewcommand\arraystretch{1}
	\begin{center}
		\begin{longtable}{lccccccccc}
			\caption{Sample of 11 sGRBs with internal X-ray plateau fits used in our analysis}
			\label{tab:1} \\
			\hline%
			GRB &${T_{90}}$  & $\Gamma$ & $z$  &$\tau$  &  $F_0$   & $a_1$ &$a_2$  & $B_p$ &$P_0$ \\
			&   (s)         &    &  &(s)  &($\times 10^{-11} erg/cm^{2}/s$ )      &    &    &  ($\times 10^{15}G$) &  ms  \\
			(1)&(2)& (3) &(4) & (5) &(6) & (7)&(8)&(9)&(10)\\
			\hline%
			\endhead%
			\hline%
			\endfoot%
			\hline%
			\endlastfoot%
			150120A&1.20&2.00&0.46&119.80 $\pm$3.55&15.72 $\pm$3.30&-0.50 $\pm$0.61&15.23 $\pm$1.73&172.63 $\pm$18.81&41.39 $\pm$4.38\\
			150301A&0.48&1.71&(0.72)&66.27 $\pm$12.81&510.54 $\pm$146.62&0.33 $\pm$0.48&2.99 $\pm$1.13&34.45 $\pm$8.30&6.14 $\pm$1.06\\
			150831A&1.15&1.64&(0.72)&99.32 $\pm$8.52&47.39 $\pm$11.57&0.80 $\pm$0.44&9.20 $\pm$10.80&76.97 $\pm$11.49&16.80 $\pm$2.17\\
			160624A&0.20&1.60&0.483&115.49 $\pm$9.73&13.75 $\pm$3.54&1.10 $\pm$0.35&5.79 $\pm$0.38&195.31 $\pm$30.02&45.98 $\pm$6.22\\
			181123B&0.26&1.99&(0.72)&262.80 $\pm$60.65&1.67 $\pm$0.58&0.70 $\pm$0.20&3.81 $\pm$3.03&141.07 $\pm$40.71&50.09 $\pm$10.43\\
			191031D&0.29&1.88&(0.72)&113.97 $\pm$103.05&30.51 $\pm$40.67&0.34 $\pm$0.15&2.55 $\pm$1.49&78.18 $\pm$87.82&18.28 $\pm$14.72\\
			200411A&0.22&1.77&(0.72)&86.86 $\pm$8.61&16.58 $\pm$3.02&0.01 $\pm$0.23&5.80 $\pm$3.52&143.52 $\pm$19.33&29.30 $\pm$3.04\\
			200907B&0.83&1.96&(0.72)&215.50 $\pm$75.38&2.34 $\pm$0.93&0.79 $\pm$0.13&4.52 $\pm$1.83&146.08 $\pm$58.70&46.97 $\pm$12.40\\
			210323A&1.12&1.72&(0.72)&108.65 $\pm$27.36&26.69 $\pm$9.59&0.36 $\pm$0.47&3.00 $\pm$1.38&91.55 $\pm$28.32&20.90 $\pm$4.59\\
			250322A&0.62&2.08&0.42&158.31 $\pm$33.62&4.08 $\pm$1.57&0.66 $\pm$0.05&4.28 $\pm$1.05&281.62 $\pm$80.74&77.62 $\pm$17.07\\
			250821A&1.52&1.62&(0.72)&373.57 $\pm$59.76&0.75 $\pm$0.22&0.40 $\pm$0.30&3.67 $\pm$0.42&163.90 $\pm$35.45&69.39 $\pm$11.53\\
		\end{longtable}
		\begin{tablenotes}
			\textbf{Note}: Redshift values $z$ are compiled from published literature and Gamma-ray Coordinates Network (GCN) circulars. For sources lacking spectroscopic or photometric redshift measurements, we adopt $z=0.72$, the mean redshift derived from all sample of measured  \textit{swift} sGRBs.
		\end{tablenotes}
	\end{center}

\newpage
\bibliography{sample631}{}
\bibliographystyle{aasjournal}

\end{document}